\pdfoutput=1

\documentclass[%
  aip,%
  jmp,%
  reprint,%
]{revtex4-1}

\usepackage{amsfonts}
\catcode`\*=11
\makeatletter
\let\equation*\@undefined
\let\endequation*\@undefined
\makeatother
\catcode`\*=12
\usepackage{amsmath}

\usepackage[protrusion=true,expansion]{microtype}

\usepackage{xargs}
\usepackage{ifthen}
\usepackage{mathtools}
\usepackage{xfrac}
\usepackage{mathrsfs}

\usepackage{graphicx}

\usepackage{xspace}
\newcommand*{\ie}{i.e.\@\xspace}
\newcommand*{\eg}{e.g.\@\xspace}
\newcommand*{\cf}{cf.\@\xspace}

\newcommand*{\vs}{vs.\@\xspace}
\makeatletter
\newcommand*{\etc}{%
    \@ifnextchar{.}%
        {etc}%
        {etc.\@\xspace}%
}
\makeatother

\let\originalleft\left
\let\originalmiddle\middle
\let\originalright\right
\renewcommand{\left}{\mathopen{}\mathclose\bgroup\originalleft}
\renewcommand{\middle}{\originalmiddle}
\renewcommand{\right}{\aftergroup\egroup\originalright}

\newcommand{\ii}{\ensuremath{\mathrm{i}}}
\newcommand{\ee}{\ensuremath{\mathrm{e}}}

\DeclarePairedDelimiter\abs{\lvert}{\rvert}
\DeclarePairedDelimiter\norm{\lVert}{\rVert}

\DeclareMathOperator{\arccsc}{arccsc}

\DeclareMathOperator{\Arg}{Arg}

\DeclarePairedDelimiterX\expval[1]{\langle}{\rangle}{#1}
\DeclarePairedDelimiterX\bra[1]{\langle}{\vert}{#1}
\DeclarePairedDelimiterX\ket[1]{\vert}{\rangle}{#1}
\DeclarePairedDelimiterX\overlap[2]{\langle}{\rangle}{#1\delimsize\vert#2}
\DeclarePairedDelimiterX\bracket[3]{\langle}{\rangle}{#1\delimsize\vert#2\delimsize\vert#3}
\DeclarePairedDelimiterX\clebschgordan[3]{\langle}{\rangle}{#1,#2\delimsize\vert#3}

\renewcommand{\vec}[1]{\ensuremath{\boldsymbol{#1}}}
\newcommand{\uvec}[1]{\ensuremath{\hat{\vec{#1}} \protect\vphantom{\vec{#1}}}}

\begin{document}

\preprint{AIP/123-QED}

\title{Certifying The Quantumness of A Generalized Coherent Control Scenario}

\author{Torsten Scholak}
\email{tscholak@chem.utoronto.ca}
\affiliation{Chemical Physics Theory Group, Department of Chemistry, 
and Center for Quantum Information and Quantum Control,
University of Toronto, Toronto, Ontario, Canada, M5S~3H6}

\author{Paul Brumer}
\affiliation{Chemical Physics Theory Group, Department of Chemistry, 
and Center for Quantum Information and Quantum Control,
University of Toronto, Toronto, Ontario, Canada, M5S~3H6}

\date{\today}

\begin{abstract}
  We consider the role of quantum mechanics in a specific coherent control
  scenario, designing a ``coherent control interferometer" as the
  essential tool that links coherent control to quantum fundamentals.
  Building upon this allows us to rigorously display the genuinely quantum nature
  of a generalized weak-field coherent control scenario
  (utilizing 1 \vs{} 2 photon excitation) via a Bell-CHSH test.
  Specifically, we propose an implementation of
  ``quantum delayed-choice'' 
  in a bichromatic alkali atom photoionization experiment.
  The experimenter can choose between two complementary situations,
  which are characterized by
  a random photoelectron spin polarization with particle-like behavior
  on the one hand, and by spin controllability and wave-like nature
  on the other. Because these two choices are conditioned coherently
  on states of the driving fields, it becomes physically unknowable, prior
  to measurement, whether there is control over the spin or not.
\end{abstract}



\maketitle

\section{Introduction}

Various coherent control scenarios, in both complex and simple systems,
utilize laser fields to coherently manipulate
atomic or molecular fragmentation processes \cite{Shapiro:2012ly}.
Two competing, fundamentally different perspectives have been advocated
to explain the control. 
In the first, within
the spirit of the Young double-slit experiment,
one interprets the probability of the
outcomes---the control target or yield---as an
intensity and phase dependent pattern resulting from the quantum-coherent
interference of mutually exclusive path alternatives
that are embodied in the laser excitation pathways
\cite{Rice:2000,Shapiro:2003,Shapiro:2012ly}.
In the second, one views the controllability as
the manifestation of the response of the system
to the superposition of phase-coherent incident laser fields.
In this approach, control is perceived as an
inherently classical phenomenon
\cite{Constantoudis:2005qo,Lima:2008vn,Ivanov:2012aa}, 
\ie{} a phenomenon that could fall under descriptions based on classical
laws of motion. Understanding which of these descriptions is preferred
is not just a matter of convenience. Rather, it has practical applications
stemming from the recognition that decoherence effects often bring a 
system to the classical limit \cite{Schlosshauer:2007sf}.
Hence, if control is indeed (at least
partially) classical, then it may well survive in the often unavoidable
decohering environments associated with realistic molecular processes.

There is ample motivation to address the issue of the role of quantum \vs{}
classical effects in coherent control. For example,
Refs.~\citenum{Franco:2006rq} and \citenum{Franco:2008fc}
analyze controlled symmetry breaking
in a field-driven quartic oscillator both quantum-mechanically
and classically and concludes that not only the basic requirements,
but also the physical origins of control, are the same in both cases.
Similarly, Ref.~\citenum{Pachon:2013ly} shows that 
environmentally assisted one photon phase control
is mainly due to the
incoherent breaking of time-reversal symmetry,
and is thereby not evidence of quantum coherent dynamics.
In addition, it is of interest to note that related concerns
regarding classical \vs{} quantum
coherence contributions have arisen within the framework of electronic
energy transfer in light-harvesting systems \cite{Miller:2012}.

Classical descriptions can offer an intuitively
appealing picture, but in the case of coherent control
they often fail quantitatively, and are hence discarded.
Therefore, since its inception, coherent control has been
regarded as fundamentally quantum
\cite{Brumer:1986uq,Shapiro:1988cr,Shapiro:2012ly},
particularly in the case of driving a system with two 
frequencies $(n \omega + m \omega)$. Here, 
reliance on an interpretation of 
the interference of quantum pathways as described above (and as
originally put forth \cite{Brumer:1986uq,Shapiro:1988cr}
by one of the authors of this paper) is
commonly accepted. But the issue of the extent to which
\emph{nontrivial} quantum features are central to phenomena such as
coherent control needs to be reconsidered in light of
developments in Bell-like tests to certify unique quantum features
\cite{Bell:1964zt,Bell:1966zr,Clauser:1969dq,Guhne:2009vn,Ionicioiu:2011fk,Peruzzo:2012mw,Kaiser:2012fk}
and recent reports that identify distinct classical mechanisms
as a possible source of control
\cite{Flach:2000ys,Sirko:2002ye,Franco:2006rq,Franco:2008fc,Lima:2008vn,Franco:2010kx,Pachon:2013ly}.

We  reiterate that there is no question that quantum mechanics is necessary to 
\emph{quantitatively} describe the outcome of coherent control scenarios.
However, quantitative agreement with a quantum description
cannot serve as a general proof
that the observed phenomenon is unambiguously quantum in nature
(since such a proof requires that all classical descriptions
and explanations must be falsified). As a consequence,
rigorously certifying the quantumness of a process and identifying its quantum
features is a challenging task that has been the subject of intense efforts in
quantum optics, quantum information and quantum foundations. Studies of this type often
take the form of proposed experimental protocols that are carefully crafted 
to close any loopholes that would prevent rigorous assertions regarding those
features of quantum mechanics that are manifest in the process. Such features
include issues such as nonlocality, entanglement, multiple-pathway interference,
sensitivity to measurement, \etc{} \cite{Greenstein:2005}. In this paper we make
the first inroads into utilizing ideas of this kind to explore the quantum
characteristics of coherent control. Specifically, we first focus on path interference
and introduce a ``coherent control interferometer''
which formalizes the relationship between
coherent control scenarios and quantum optics approaches to the
fundamentals of quantum mechanics \cite{Greenstein:2005,Haroche:2006}. As a 
particular case we concentrate
the analysis on phase-coherent control over
the spin polarization of an electron
that has been emitted in an
interfering $(\omega + 2 \omega)$ photoionization process.

We then design a \emph{generalized} coherent control scenario
in which measurements can be performed
whose outcomes will have certifiably nonclassical statistics
and that strongly support the analogy between coherent control
and the quantum interference of paths. We show that this setup
can be used to probe wave-particle complementarity
and to implement ``quantum delayed-choice''
\cite{Ionicioiu:2011fk,Peruzzo:2012mw,Kaiser:2012fk};
it becomes unforeseeable prior to measurement
whether the spin polarization statistics are wave- or particle-like,
that is, whether there is control or not.
This hallmark of quantum interference is key for
violating a Bell inequality
that serves as a rigorous experimental test of local realism
and thereby distinguishes nonclassical from classical statistics,
entirely on the basis of observed statistical data
\cite{Bell:1964zt,Bell:1966zr,Clauser:1969dq,Guhne:2009vn}.

Note the fact that we deal with a generalization of traditional
coherent control scenarios indicates the continuing need to identify
methods of certifying the quantumness of traditional coherent control
scenarios. Such studies are underway and our expectation is that the coherent control 
interferometer introduced in this paper will serve as a central tool for
such studies.

\section{Control Scenario}

\subsection{Coherent Control Interferometer}
\label{ccinterferometer}

\begin{figure}
  (a) \hspace{.1cm} \raisebox{-15.57982pt}{\includegraphics{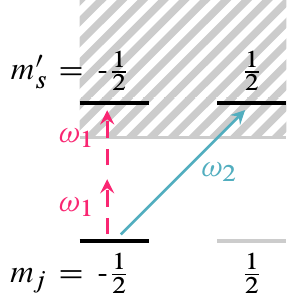}}%
  \hspace{.5cm}%
  (b) \hspace{.1cm} \raisebox{-15.57982pt}{\includegraphics{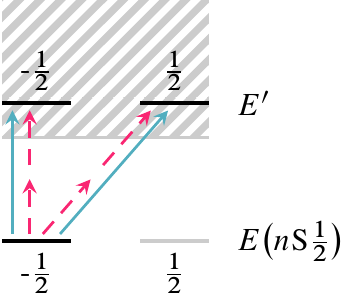}}%
  \caption{%
    (color online)
    Proposed coherent control interferometer (CCI).
    (a) In the open configuration $o$,
    single-photon ionization (solid blue) only undergoes
    a $\sigma$-transition with $\abs[\big]{m_{j} - m_{s}'} = 1$,
    whereas two-photon ionization (dashed red)
    leaves the electron spin unchanged.
    The processes are distinguishable from one another
    due to their different outcomes.
    (b) This is not the case in the closed setup $c$,
    for which the transition amplitudes between
    all spin combinations are nonzero.%
  }%
  \label{fig:cci}%
\end{figure}

Let a heavy alkali atom be ionized
by weak coherent ($\omega + 2\omega$) radiation
\cite{Yin:1992fu,Yin:1993fk,Wang:2001pi}.
For purposes of simplification (rather than physical necessity),
we assume a tight confinement that fully suppresses
decoherence due to recoil
\footnote{\cf{}
Lamb-Dicke regime \cite{Grimm:2000tg,Eschner:2003kl}.
Interference can of course still be observed with some weak recoil.}.
The control target is
the laboratory $\uvec{z}$-axis projection $m_{s}'$
of the photoelectron's spin in the continuum
\cite{Fano:1969tg,Baum:1970hc}.
We consider the case
where the continuum state $\ket*{\vec{K}', m_{s}' {}^{+}}$
of the electron at energy $E' = \left(\hbar K'\right) {}^{2} / 2 m_{e}$
can be reached from the atomic ground state
$\ket*{n \mathrm{S} \frac{1}{2}, m_{j}}$
mainly by two pathways \cite{Taylor:1972nx}:
(i) absorption of $1$ photon of energy 
$\hbar \omega_{2} = E' - E \left(n \mathrm{S} \frac{1}{2}\right)$
or (ii) of $2$ photons each with energy 
$\hbar \omega_{1} = \hbar \omega_{2} / 2$.
Here, $\vec{K}'$ refers to the electron's
asymptotic outgoing wavevector \cite{Taylor:1972nx},
$n$ is the ground state's principle quantum number \cite{Bethe:2008oq},
and the projection of the total angular momentum $j$
onto the $\uvec{z}$-axis is denoted $m_{j}$.
This setup implements a ``coherent control
interferometer" (CCI), Fig.~\ref{fig:cci},
an analog of a Mach-Zehnder interferometer (MZI).
The ground state is spin-$\frac{1}{2}$
and constitutes the CCI's two input ports
whereas the final projections $m_{s}' = \pm \frac{1}{2}$
are identified with the two output ports,
the measurement statistics $p_{\pm}$ of which
are interpreted as the interference patterns.
Henceforth, we restrict attention to one input port only,
$m_{j} = - \frac{1}{2}$.

The incoming laser modes $l$
are to be populated with coherent states of light,
$\ket{l} \equiv \ket*{\left(\alpha \vec{k} \uvec{\varepsilon}\right)_{l}}$,
since these are eigenstates of the photon annihilation operator
and thus correlations with matter due to photon absorption
do not occur \cite{Gong:2010dq}.
The a priori probabilities of the ionization pathways
can be adjusted by means of
the amplitude moduli $\abs*{\alpha_l}$,
the directions of incidence $\vec{k}_{l}$,
and the polarizations $\uvec{\varepsilon}_{l}$.
This is because (as described in the appendix),
in the long-time limit, the change of the electron's spin
is described in terms of an infinite series of
well-known, polarization-dependent transition matrix amplitudes
$t_{i} \bigl(m_{s}', m_{j}^{\vphantom{\prime}}\bigr)$,
$i = 1, 2, \ldots$,
multiplied by corresponding powers
of the field amplitudes $\alpha_{l}$
\cite{Cohen-Tannoudji:1998kl,Fermi:1930qa,Seaton:1951mi,Cooper:1962nx,Burgess:1960ij,Bebb:1966fv,Lambropoulos:1976dz}.
In the present case,
only the first two transition amplitudes are relevant
\cite{Yin:1992fu,Yin:1993fk};
$t_{1}$ resolves the $1$-photon process and 
connects only to $\mathrm{P}$ continuum states,
whereas $t_{2}$  accounts for $2$-photon ionization
and accesses $\mathrm{S}$ and $\mathrm{D}$ orbitals.
The amplitudes contain
radial integrals, $D_{1}$ and $D_{2}$,
that serve here as
complex empirical parameters.
There exist measurement schemes
for which the ionization pathways become absolutely distinguishable.
Therefore,
in what follows,
spin statistics are conditioned on
successful detection of the electron
in the channel $\uvec{K}' = \bigl(\sqrt{2} \, \uvec{x} - \uvec{z}\bigr) / \sqrt{3}$.
This both eliminates the need for taking into account
the efficiency of detection and
does not provide information about the path 
taken from the initial to final state of the atom.

\begin{figure}
  \includegraphics{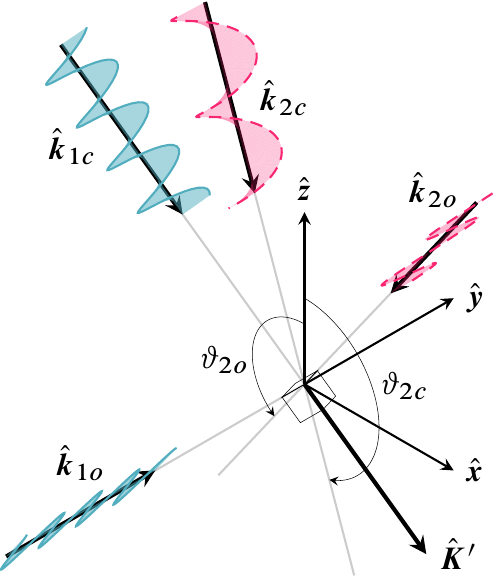}%
  \caption{%
    (color online)
    Sample scattering geometry with incident light wavevectors
    $\uvec{k}_{1 o} = \uvec{y}$, $\uvec{k}_{2 o}$
    with projection
    $\uvec{k}_{2 o} \cdot \uvec{z} = \cos \vartheta_{2 o} = - \sqrt{2 / 3}$,
    $\uvec{k}_{1 c} = \uvec{K}'$,
    and $\uvec{k}_{2 c}$
    with $\cos \vartheta_{2 c} = - 2 \sqrt{2} / 3$.
    The emitted photoelectron is detected
    in the direction $\uvec{K}'$.
    The alkali atom is located at the origin.
    The polarization of the incoming radiation
    is indicated in color (shades of grey).
    The $o$ configuration uses linear polarization,
    whereas the $c$ setup employs elliptically polarized light.%
  }%
  \label{fig:geometry}%
\end{figure}

We argue that
quantum interference in the CCI can be exposed
in the same manner as it can be displayed
in the Young double-slit \cite{Scully:1991kl,Itano:1998hc},
a MZI,
or a Ramsey interferometer \cite{Englert:1996bs,Bertet:2001ij}
for single photons, electrons, \etc.
The latter undertaking entails the observation of
complementarity between the wave and particle property,
which is demonstrated by
the experimenter's choice between two measurement statistics:
particle and wave statistics.
In the case of the former,
interference is completely absent.
In principle, if not in practice,
knowledge about the path%
---whether the photon has taken the first rather than the second slit,
or, correspondingly,
whether the atom has been ionized in a $1$- or a $2$-photon
process---is available by measurement.
When it allows for tracing of paths, 
an interferometer is called open.
If, in contrast, interference is displayed,
then the statistics are wave-like.
Full path knowledge cannot be acquired,
not even in principle.
This results from operating the interferometer
in the so-called closed configuration.
Below, the subscripts
$o$ and $c$ are used to denote these configurations.

The ionization scenario has been selected
because it is conducive to engineering
both the $o$ and $c$ cases,
Fig.~\ref{fig:geometry}.

\subsubsection{The Open Interferometer Configuration}

For the open configuration, two linearly polarized modes,
$\left(\vec{k} \uvec{\varepsilon}\right)_{1 o}$
and $\left(\vec{k} \uvec{\varepsilon}\right)_{2 o}$,
are chosen, for example, to be populated according to
$\uvec{\varepsilon}_{1 o} = - \uvec{k}_{2 o} =
\bigl(\uvec{x} + \sqrt{2} \, \uvec{z}\bigr) / \sqrt{3}$
and $\uvec{\varepsilon}_{2 o} = \uvec{k}_{1 o} = \uvec{y}$
in a field state denoted $\ket*{1o, 2o}$.
The vectors $\uvec{K}'$, $\uvec{k}_{2 o}$, and $\uvec{k}_{1 o}$
come as a right-handed orthogonal triad
that is inclined at an angle $\pi - \arctan (\sqrt{2})$
away from the positive $\uvec{z}$-axis.
These mode settings imply $t_{1 o} \propto \delta_{m_{s}', - m_{j}}$
and $t_{2 o} \propto \delta_{m_{s}', m_{j}}$
and, as a result, absolute path knowledge;
the direction in which the spin is detected
reveals the path taken through the CCI,
\cf{} Figs.~\ref{fig:cci} (a) and \ref{fig:geometry}.
To realize an unbiased interferometer,
the moduli of the complex amplitudes
are to be adjusted as follows:
\begin{subequations}
  \begin{align}
    \abs*{\alpha_{1 o}} & = \frac{6 \sqrt{\pi} \lambda \ee^{\ii \kappa_{1 o}}}{\hbar g_{k_{1}} \left[- D_{1} \left(E' \mathrm{P} \tfrac{1}{2}, n \mathrm{S} \tfrac{1}{2}\right) + D_{1} \left(E' \mathrm{P} \tfrac{3}{2}, n \mathrm{S} \tfrac{1}{2}\right)\right]},
    \label{eq:alpha1o} \\
    \abs*{\alpha_{2 o}}^2 & = \frac{90 \sqrt{\pi} \lambda \ee^{\ii \kappa_{2 o}}}{\ii \hbar_{\vphantom{k_{2}}}^{2} g_{k_{2}}^{2}} \bigl[- 5 D_{2} \left(E' \mathrm{S} \tfrac{1}{2}, \mathrm{P} \tfrac{1}{2}, n \mathrm{S} \tfrac{1}{2}\right) \nonumber \\ \MoveEqLeft - 10 D_{2} \left(E' \mathrm{S} \tfrac{1}{2}, \mathrm{P} \tfrac{3}{2}, n \mathrm{S} \tfrac{1}{2}\right) + 5 D_{2} \left(E' \mathrm{D} \tfrac{3}{2}, \mathrm{P} \tfrac{1}{2}, n \mathrm{S} \tfrac{1}{2}\right) \nonumber \\ \MoveEqLeft + D_{2} \left(E' \mathrm{D} \tfrac{3}{2}, \mathrm{P} \tfrac{3}{2}, n \mathrm{S} \tfrac{1}{2}\right) + 9 D_{2} \left(E' \mathrm{D} \tfrac{5}{2}, \mathrm{P} \tfrac{3}{2}, n \mathrm{S} \tfrac{1}{2}\right)\bigr]^{-1}.
    \label{eq:alpha2o} 
  \end{align}
  \label{eq:alphao}%
\end{subequations}
The Eqs.~\eqref{eq:alphao}
define conditions on the phases
$\kappa_{1 o}$ and $\kappa_{2 o}$
that ensure that the right hand sides
of Eq.~\eqref{eq:alpha1o} and \eqref{eq:alpha2o}
are real and nonnegative.
$\kappa_{1 o}$ and $\kappa_{2 o}$
therefore depend on the complex phases
of the integrals $D_{1}$ and $D_{2}$
and will henceforth be referred to as
material phases.
The arguments of $D_{1}$ and $D_{2}$
refer to the respective photoionization channels. For example,
$D_{2} \left(E' \mathrm{D} \tfrac{5}{2}, \mathrm{P} \tfrac{3}{2}, n \mathrm{S} \tfrac{1}{2}\right)$
is the radial matrix element
of the process leading from the ground state
$\ket[\big]{n \mathrm{S} \tfrac{1}{2}, m_{j}^{\vphantom{\prime}}}$
via intermediate $\mathrm{P} \tfrac{3}{2}$ orbitals
to the continuum state
$\ket[\big]{E' \mathrm{D} \tfrac{5}{2}, m_{j}' {}^{+}}$.
The field's phases,
$\phi_{1 o} = \Arg \alpha_{1 o}$ and $\phi_{2 o} = \Arg \alpha_{2 o}$,
can still be chosen freely. In Eqs.~\eqref{eq:alphao},
$\lambda$ is a positive, constant scaling factor.
The precise value of $\lambda$ is irrelevant,
since the probability amplitudes are
eventually conditioned on successful detection
of the photoelectron in the channel $\uvec{K}'$.
For the choices above,
$\abs*{t_{1 o}^{\vphantom{2}} \alpha_{1 o}^{\vphantom{2}}} = \abs*{t_{2 o}^{\vphantom{2}} \alpha_{2 o}^{2}}$.
We have thus established that, for the initial state
$\ket*{n \mathrm{S} \tfrac{1}{2}, \text{-} \tfrac{1}{2}} \otimes \ket*{1 o, 2 o}$,
the final state of the ionized atom in the channel $\vec{K}'$ is
$\ket*{f_{o}} \otimes \ket*{1 o, 2 o}$, where
\begin{align}
 \ket*{f_{o}} & = \frac{\ee^{\ii \delta_{o}}}{\sqrt{2}} \, \left(\ket*{\vec{K}', \tfrac{1}{2} {}^{+}} + \ee^{- \ii \theta_{o}} \ket*{\vec{K}', \text{-} \tfrac{1}{2} {}^{+}}\right)
 \label{eq:fo}
\end{align}
with $\delta_{o} = \phi_{1 o} + \kappa_{1 o}$
and $\theta_{o} = \phi_{1 o} - 2 \phi_{2 o} + \kappa_{1 o} - \kappa_{2 o}$.
The state $\ket*{f_{o}}$ describes full particle-like statistics;
projected onto the $\uvec{z}$ axis,
both spin orientations $m_{s}' = \pm \frac{1}{2}$ are always equally likely,
regardless of the lasers' phase difference
$\phi_{1 o} - 2 \phi_{2 o}$.
Thus there is no phase-coherent control
and the fringe visibility vanishes on both output ports.

\subsubsection{The Closed Interferometer Configuration}

As for the closed configuration $c$,
it can be achieved by
populating two other modes,
$\left(\vec{k} \uvec{\varepsilon}\right)_{1 c}$ and
$\left(\vec{k} \uvec{\varepsilon}\right)_{2 c}$
(see Fig.~\ref{fig:geometry}). For example,
the one-photon absorption field $1 c$
is chosen circularly polarized,
$\sqrt{14} \, \uvec{\varepsilon}_{1 c} = \sqrt{5 \smash{/} 3 + \sqrt{2}} \, \bigl(\uvec{x} + \sqrt{2} \, \uvec{z}\bigr) - \ii \sqrt{9 - 3 \sqrt{2}} \, \uvec{y}$
with $\uvec{k}_{1 c} = \uvec{K}'$,
whereas the two-photon absorption field $2 c$
is chosen elliptically polarized,
$\sqrt{2} \, \uvec{\varepsilon}_{2 c} = \bigl(2 \sqrt{2} \, \uvec{x} + \uvec{z}\bigr) / 3 + \ii \uvec{y}$
as well as
$\uvec{k}_{2 c} = \bigl(\uvec{x} - 2 \sqrt{2} \, \uvec{z}\bigr) / 3$,
making an angle of $\pi - \arccsc (3)$
with the $\uvec{z}$-axis.
The resultant field state is denoted $\ket*{1c, 2c}$.
Only $t_{2} \left(\tfrac{1}{2}, \tfrac{1}{2}\right)$
and $t_{2} \left(\text{-} \tfrac{1}{2}, \text{-} \tfrac{1}{2}\right)$
contain $\mathrm{S}$-wave $D_{2}$ factors.
They are fully suppressed in the chosen field configuration.
(It turns out that simultaneous unbiased interference
for both input ports of the CCI, $m_{j} = \pm \frac{1}{2}$,
cannot be achieved. That is why the above settings are tailored specifically
to balance the interference from the input port $m_{j} = - \frac{1}{2}$).
With these, we find
$t_{1 c} \left(\tfrac{1}{2}, \text{-} \tfrac{1}{2}\right)
= t_{1 c} \left(\text{-} \tfrac{1}{2}, \text{-} \tfrac{1}{2}\right)$.
Although it would be most desirable to also have
$t_{2 c} \left(\tfrac{1}{2}, \text{-} \tfrac{1}{2}\right)
= - t_{2 c} \left(\text{-} \tfrac{1}{2}, \text{-} \tfrac{1}{2}\right)$,
the latter amplitudes are somewhat biased
and have generally a mutual phase shift different from $\pi$.
This is the case unless one can select an energy $E'$ for which
the transition to $j' = \frac{5}{2}$
is negligible compared to $j' = \frac{3}{2}$;
indeed,
for $D_{2} \left(E' \mathrm{D} \tfrac{5}{2}, \mathrm{P} \tfrac{3}{2}, n \mathrm{S} \tfrac{1}{2}\right) = 0$,
the $t_{2 c}$ are as desired.
If this were the case 
the amplitudes would be set equal in absolute value to
\begin{subequations}
  \begin{align}
    \abs*{\alpha_{1 c}} & = \sqrt{\tfrac{7}{2 \left(3 - \sqrt{2}\right)}} \, \abs*{\alpha_{1 o}}, \\
    \abs*{\alpha_{2 c}}^{2} & = \frac{180 \sqrt{\pi} \lambda \ee^{\ii \kappa_{2 c}}}{\ii \bigl(\sqrt{2} + 2\bigr) \hbar^{2} g_{k_{2}}^{2}} \bigl[5 D_{2} \left(E' \mathrm{D} \tfrac{3}{2}, \mathrm{P} \tfrac{1}{2}, n \mathrm{S} \tfrac{1}{2}\right) \nonumber \\ \MoveEqLeft + D_{2} \left(E' \mathrm{D} \tfrac{3}{2}, \mathrm{P} \tfrac{3}{2}, n \mathrm{S} \tfrac{1}{2}\right)\bigr]^{-1}
    \label{eq:alpha2c}
  \end{align}
  \label{eq:alphac}%
\end{subequations}
in order to achieve unbiased interference.
We choose the amplitudes \eqref{eq:alphac} even for cases
where $D_{2} \left(E' \mathrm{D} \tfrac{5}{2}, \mathrm{P} \tfrac{3}{2}, n \mathrm{S} \tfrac{1}{2}\right) \neq 0$.
The phases $\phi_{1 c} = \Arg \alpha_{1 o}$
and $\phi_{2 c} = \Arg \alpha_{2 o}$
are free parameters,
whereas $\kappa_{2 c}$
is another constant material phase
that is implicitly defined by Eq.~\eqref{eq:alpha2c}.
The phases picked up by the one-photon processes
are identical in the configurations $o$ and $c$,
therefore $\kappa_{1 c} = \kappa_{1 o}$.
Consider now initialization from the state
$\ket*{n \mathrm{S} \tfrac{1}{2}, \text{-} \tfrac{1}{2}} \otimes \ket*{1 c, 2 c}$.
For successful detection in the channel $\vec{K}'$,
the final state becomes
$\ket*{f_{c}} \otimes \ket*{1 c, 2 c}$, where
\begin{align}
  \ket*{f_{c}} & = \frac{\ee^{\ii \delta_{c}}}{N_{f_{c}}} \, \Bigl[\left(\ii \sin \tfrac{\phi_{c} + \theta_{c}}{2} - \ee^{- \ii \phi_{c} / 2} d_{2}\right) \ket*{\vec{K}', \tfrac{1}{2} {}^{+}} \nonumber \\ \MoveEqLeft + \left(\cos \tfrac{\phi_{c} + \theta_{c}}{2} + \left(\tfrac{7}{2} - \tfrac{5}{\sqrt{2}}\right) \ee^{- \ii \phi_{c} / 2} d_{2}\right) \ket*{\vec{K}', \text{-} \tfrac{1}{2} {}^{+}}\Bigr]
  \label{eq:fc}
\end{align}
with the global phase
$\delta_{c} = \left(\phi_{1 c} + 2 \phi_{2 c} + \kappa_{1 c} + \kappa_{2 c}\right) / 2$,
the interferometric phase difference $\phi_{c} = \phi_{1 c} - 2 \phi_{2 c}$,
the material phase shift $\theta_{c} = \kappa_{1 c} - \kappa_{2 c}$,
the normalization $N_{f_{c}}$, and with
\begin{align}
  d_{2} & = \frac{- 3 \ii \, \ee^{- \ii \left(\kappa_{1 c} + \kappa_{2 c}\right) / 2} D_{2} \left(E' \mathrm{D} \tfrac{5}{2}, \mathrm{P} \tfrac{3}{2}, n \mathrm{S} \tfrac{1}{2}\right)}{\abs*{5 D_{2} \left(E' \mathrm{D} \tfrac{3}{2}, \mathrm{P} \tfrac{1}{2}, n \mathrm{S} \tfrac{1}{2}\right) + D_{2} \left(E' \mathrm{D} \tfrac{3}{2}, \mathrm{P} \tfrac{3}{2}, n \mathrm{S} \tfrac{1}{2}\right)}}.
\end{align}
The state $\ket*{f_{c}}$ is wave-like;
coherent control of the electron's spin is possible
by virtue of the phase $\phi_{c}$.
The interference contrast
is maximal for $D_{2} \left(E' \mathrm{D} \tfrac{5}{2}, \mathrm{P} \tfrac{3}{2}, n \mathrm{S} \tfrac{1}{2}\right) = 0$,
in which case also $d_{2} = 0$ and
$\abs*{t_{1 c} \left(m_{s}', \text{-} \tfrac{1}{2}\right) \, \alpha_{1 c}^{\vphantom{2}}}
= \abs*{t_{2 c} \left(m_{s}', \text{-} \tfrac{1}{2}\right) \, \alpha_{2 c}^{2}}$.

\subsection{Complementarity}

Complementarity in quantum interference demands
that the experimental situation determines
whether one detects the statistics of a wave or a particle
(or even a blend of the two) \cite{Bohr:1984fu,Englert:1998kl}.
In classical physics,
these concepts are mutually exclusive,
and the object passing the interferometer
cannot subscribe to either of them ad libitum.
In order to stress this fundamental difference,
Wheeler proposed the ``delayed-choice'' experiment
\cite{Wheeler:1984cq}
that severs any causal link between
the object and the interferometer until it has entered it.
If we wanted to realize
Wheeler's gedanken experiment with coherent control,
we would have to postpone and randomize
the choice between $o$ and $c$
such that the spin cannot ``know'' beforehand
which property to display.
The proposed CCI does not, however, literally allow this,
because randomization between
$\ket*{1 o, 2 o}$ and $\ket*{1 c, 2 c}$
would choose the ``control" or ``no control" in advance.
(Indeed this is the central stumbling block to considering
features of complementarity in coherent control scenarios).
Notwithstanding, instead of flipping a coin between $o$ and $c$,
we can design a generalized coherent control scenario where one prepares
a coherent superposition 
$\ket*{i} = \ket*{n \mathrm{S} \tfrac{1}{2}, \text{-} \tfrac{1}{2}} \otimes \left(\ket*{1 o, 2 o} + \ket*{1 c, 2 c}\right) / N_{i}$
that gives the final state
\begin{align}
  \ket*{f} & = \frac{1}{N_{f}} \, \left(\ket*{f_{o}} \otimes \ket*{1 o, 2 o}
  + N_{f_{c}} \ket*{f_{c}} \otimes \ket*{1 c, 2 c}\right)
  \label{eq:f}
\end{align}
with $N_{i}$, $N_{f}$ being normalization factors.
Since the states $\ket*{1 o, 2 o}$ and $\ket*{1 c, 2 c}$
are each direct products of two coherent states in,
in total, four mutally orthogonal modes,
the initial state $\ket*{i}$ features nonclassical correlations
of the GHZ type \cite{Sanders:1992zr,Jeong:2006kx,Greenberger:1990oq,Guhne:2009vn}
the creation of which
requires a special nonlinear MZI
\footnote{Specifically,
entangled coherent states are produced
from a single coherent light source (\eg{}, a laser)
by letting the light interfere with itself
in a MZI in which one arm
has been replaced with a strong
Kerr nonlinearity \cite{Sanders:1992zr}.
Furthermore, in order to prepare the initial state $\ket*{i}$,
it is not enough to create entangled light
with a single frequency $\omega$.
This is because,
compared to the two-photon ionization process,
the one-photon process needs
light with a doubled frequency $2 \omega$.
The next step in the preparation of $\ket*{i}$
is therefore second harmonic generation
which is also a nonlinear optical process.
Further standard steps are used
to balance the intensities of the beams
and to adjust their phases and polarizations.
}.
By contrast,
the resultant final state $\ket*{f}$
of the total system describes
entanglement between the radiation field and the electron spin.
It is a coherent blend of
particle- and wave-like spin polarization statistics that is
conditioned on the state of the radiation field.
As such,
the field-matter entanglement
plus the superposition state character
of Eq.~\eqref{eq:f} allows for the
delayed-choice determination of whether
the system is in $\ket*{f_{o}}$ or $\ket*{f_{c}}$
conditioned on the measurement of the field.

\subsection{Bell Test}

We still owe a test
that can certify the nonclassicality
of an actual experimental realization of the CCI.
Such test is mandatory,
since only if the correlations between
the radiation field and the electron spin are
of nonclassical nature,
we can truly delay the choice of
whether the spin statistics
are that of a particle or a wave.
The statistics must be unknowable
prior to the measurement of the field.
A test for nonclassicality
is provided by a certain family of Bell-CHSH inequalities
that are derived from the concept of local realism
\cite{Bell:1964zt,Bell:1966zr,Clauser:1969dq,Guhne:2009vn}.
Adapting the approach of Ref.~\citenum{Park:2012fk},
we define two parametric dichotomic measurement operators:
The operator
$\Gamma \left(\zeta\right) = R \left(\zeta\right) {}^{\dagger} \sigma_{z} R \left(\zeta\right)$
with the rotation $R \left(\zeta\right) = \exp (\zeta \sigma_{+} - \zeta^{*} \sigma_{-})$
measures the spin asymmetry in the direction of space
defined by the complex number $\zeta$.
Here, the Pauli operators $\sigma_{z}$, $\sigma_{\pm}$ are defined
with respect to the basis
$\left\{\ket*{\vec{K}^{\smash{\prime}}, \frac{1}{2} {}^{+}},
\ket*{\vec{K}^{\smash{\prime}}, \text{-} \frac{1}{2} {}^{+}}\right\}$.
The observable
$A \left(\vec{\beta}\right) = \bigotimes_{l} F_{l} \left(\beta_{l}\right) {}^{\dagger} \cdot \left(2 \ket*{\mathrm{vac}} \bra*{\mathrm{vac}} - 1\right) \cdot \bigotimes_{l^{\smash{\prime}}} F_{l^{\smash{\prime}}} \left(\beta_{l^{\smash{\prime}}}\right)$
with the displacements
$F_{l} \left(\beta_{l}\right) = \exp \bigl(\beta_{l}^{\vphantom{*}} a_{l}^{\dagger} - \beta_{l}^{*} a_{l}^{\vphantom{\dagger}}\bigr)$
allows for a joint photon threshold measurement
over all modes $l = 1 o$, $1 c$, $2 o$, and $2 c$,
where $\ket*{\mathrm{vac}}$  denotes the vacuum state,
in all modes, of the radiation field.

\begin{figure}
  \includegraphics{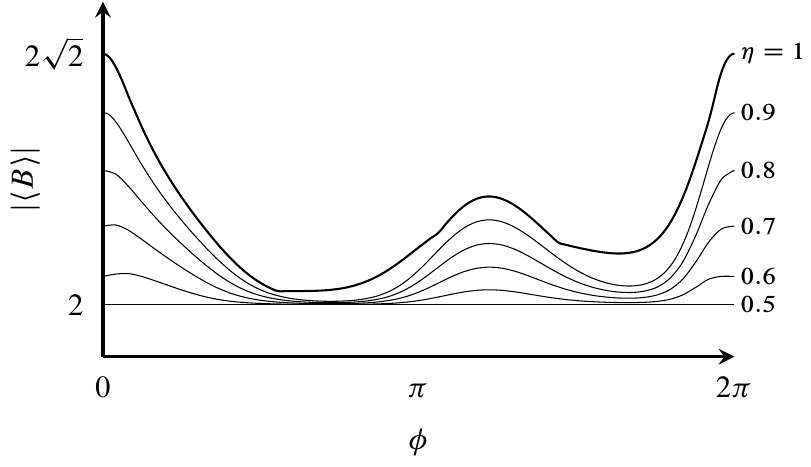}%
  \caption{%
    Numerically optimized Bell-CHSH inequality test
    for $\ket*{f}$ as a function of the phase $\phi$.
    The Bell inequality can be violated
    as long as the detection efficiency $\eta$
    is larger than $0.5$.
    For ideal photodetectors ($\eta = 1$),
    the violation can be saturated,
    $\abs*{\expval*{B}} = 2 \sqrt{2}$ at $\phi = 0$;
    the smallest violation is $2.044$. Here, for simplicity, 
     $\phi = - \phi_c - \theta_c  =  \pi/2 - \theta_o$.
  }%
  \label{fig:violation}%
\end{figure}

As described in the appendix D,
we have calculated numerically the maximally achievable violation
of the Bell-CHSH inequality 
\begin{multline}
  \abs*{\expval*{B}} \equiv \bigl\lvert\expval*{
    \Gamma \left(\zeta^{\vphantom{\prime}}\right) \otimes A \left(\vec{\beta}^{\vphantom{\prime}}\right)}
    + \expval*{\Gamma \left(\zeta'\right) \otimes A \left(\vec{\beta}^{\vphantom{\prime}}\right)} \\
    + \expval*{\Gamma \left(\zeta^{\vphantom{\prime}}\right) \otimes A \left(\vec{\beta}'\right)}
    - \expval*{\Gamma \left(\zeta'\right) \otimes A \left(\vec{\beta}'\right)
  }\bigr\rvert \le 2
  \label{eq:CHSH}
\end{multline}
by measurements on a system in the state $\ket*{f}$
as a function of $\phi$
for different photon detection efficiencies $\eta$.
The primed parameters $\zeta'$, $\vec{\beta}'$ define an additional set of observables.
To incorporate limited efficiency,
each mode is attenuated by
a beam splitter with transmissivity $\sqrt{\eta}$
before ideal detection \cite{Park:2012fk,Yuen:1980fk}.
The statistical error in the
photoelectron polarization asymmetry measurements
is not addressed at this time \cite{Osterwalder:2006vn}.
Significantly, as is evident from Fig.~\ref{fig:violation},
Bell's inequality can be violated to various degrees
for all phases $\phi$, 
where $\phi = - \phi_c - \theta_c  =  \pi/2 - \theta_o$, if $\eta > 0.5$. 
For $\phi = 0$ and $\eta = 1$, $\abs*{\expval*{B}}$ can even reach
the Tsirelson bound $2 \sqrt{2}$, the maximum allowed by quantum mechanics
\cite{Tsirelson:1980fk,Barrett:2005uq}. For every $\phi$,
the maximum violation is independent of
the absolute field amplitudes $\abs*{\alpha_{l}}$,
since the parameters $\abs*{\beta_{l}^{\vphantom{\prime}}}$
and $\abs*{\beta_{l}'}$ can always be adapted accordingly.
The optimized complex phases of the 
$\vec{\beta}$ and $\vec{\beta}'$ vary piecewise continously with $\phi$,
\footnote{At the kinks of the $(\eta = 1)$ graph,
the global maximum
switches between two different
locally optimal solutions
one of which is less robust to inefficient detection
than the other.
In case of the remaining graphs for which $\eta \le 0.9$,
the global optimum is provided
by a variation of the more robust solution
and kinks do thus not appear.}
and they are independent of $\abs*{\alpha_{l}}$.
So too are the optimal choices for $\zeta$ and $\zeta'$.
Additional numerical experiments have shown
that successful Bell test violation does not critically depend
on how small $d_{2}$ is;
near to maximum violations can still be achieved
if $\abs*{d_{2}}$ is of the order of $1$,
see the appendix.

We can now summarize as follows:
The coherent control interferometer configurations,
$o$ and $c$, create experimental conditions under which
a particle and a wave property, respectively, are displayed.
In the specific scenario that starts with
the initial state $\ket*{i}$ (above Eq.~\eqref{eq:f}),
the coherent control dynamics transform
the initial GHZ entanglement between the modes of the field
into all-encompassing nonclassical EPR correlations between
spin and radiation.
If a Bell violation is found (for a range of different $\phi$),
then it is unknowable
whether the measured spin statistics
are that of a wave or a particle
and thus whether control is possible or not.
The outcome is physically guaranteed to be random
and the experiment implements ``quantum delayed-choice''
\cite{Ionicioiu:2011fk,Tang:2012kx,Peruzzo:2012mw,Kaiser:2012fk,Roy:2012uq,Auccaise:2012kx,Stassi:2012fk}.

\section{Discussion}

It is advantageous to make explicit the difference between
the traditional understanding of the quantum character of coherent control
and the view considered in this paper.
It is commonly agreed that, in the perturbative limit, the properties of
the driving fields and of the atomic or molecular system 
independently contribute to the response
\cite{Shapiro:2012ly,Ivanov:2012aa}. 
In multi-color laser-induced coherent 
control, for example, control via phase sensitivity emerges solely from cross terms
containing products of the $n \omega$ and $m \omega$ driving field amplitudes. 
This basic mechanism is common to					       
all classical and quantum descriptions
of weak-field $(n \omega + m \omega)$
coherent control \cite{Shapiro:2012ly,Franco:2006rq,Franco:2010kx}.
It has led research to be primarily concerned with the \emph{magnitude}
of the terms giving rise to phase dependence
and, in view of quantum-classical correspondence,
whether this magnitude depends crucially
on quantum effects such as tunneling \cite{Ivanov:2012aa}
or conservation of parity \cite{Franco:2006rq}.
In other words, the divide between quantum and classical control
has been defined in terms of quantitative indicators only,
indicators which require a comparison of classical and quantum 
model calculations as a means
of assessing the importance of quantum features.

By contrast, we focus here on a fundamentally qualitative feature: the non-local
character of the proposed light-matter scenario. The resultant physics 
in the generalized coherent control scenario described here 
rules out any local realistic (classical) theory,
because controllability is conditioned on
nonclassical correlations between matter and radiation.
In the experiment, this is ascertainable a posteriori by virtue of a 
Bell-CHSH test. Inctoducing such a test has equired, however, that we go
beyond traditional coherent control scenarios to introduce and examine a
generalized scenario allowing a Bell-CHSH test. Studies refocusing attention
on traditional coherent control scenarios are underway.

\section*{Acknowledgements}

The authors benefited from helpful remarks by
Ari Mizel and Aephraim Steinberg on an earlier draft of this manuscript.
T.S.~enjoyed stimulating discussions with Klaus M\o{}lmer, Arjendu Pattanayak, and Joel Yuen. 
Financial support from NSERC Canada is gratefully 
acknowledged.

\appendix

\section{Scattering Matrix}

We examine here the interaction between
a single heavy alkali atom and
ionizing bichromatic radiation.
As explained in the main text
atomic motion is not considered,
because the atom is assumed to be tightly trapped.
Neglecting hyperfine structure,
we specify the hydrogen-like electronic bound states in the
standard spectroscopic notation as
$\ket*{n l j, m_{j}}$ with $n$ the principle quantum number
and $E \left(n l j\right)$ the energy.
The continuum states are denoted either by
$\ket*{E l j, m_{j} {}^{+}}$ with $l$ the orbital
and $j$ the total angular momentum
or by $\ket*{\vec{K}, m_{s} {}^{+}}$,
where $\vec{K}$ refers to the electron's
asymptotic outgoing wavevector \cite{Taylor:1972nx}.
In the basis of bound and continuum states, the atomic
Hamiltonian $H_{\mathcal{A}}$ is diagonal.
The bound and continuum solutions
$R_{n l j}$, $R_{E l j}$
of the radial Schr\"{o}dinger equation depend on both
the orbital angular momentum $l$
and the total angular momentum $j$;
$m_{j}$ denotes its projection onto the $\uvec{z}$-axis
in the laboratory frame.
Using Clebsch-Gordan coefficients, we write
\begin{align}
  \overlap*{\vec{x}}{n l j, m_{j}} & = \sum_{m_{l} m_{s}} \, \clebschgordan*{l m_{l}}{\tfrac{1}{2} m_{s}}{j m_{j}} \, R_{n l j} \left(x\right) \, Y_{l m_{l}} \left(\uvec{x}\right) \ket*{m_{s}},
\end{align}
where $m_{l}$ is the orbital angular momentum's projection
on the $\uvec{z}$-axis
and $\ket{m_{s}}$ describes the
electron's spin projected onto the same axis.
An analogous
expansion exists for the continuum wave functions
$\overlap*{\vec{x}}{E l j, m_{j} {}^{+}}$.
The quantized radiation field,
with canonical Hamiltonian $H_{\mathcal{R}}$,
constitutes an auxiliary degree of freedom $\mathcal{R}$.
We treat the light-matter interaction $V$
within the electric dipole approximation,
\begin{align}
  V & = e \sum_{\vec{k} \uvec{\varepsilon}} \hbar g_{k} \left(\ii \, a_{\vec{k} \uvec{\varepsilon}} \, \vec{x} \cdot \uvec{\varepsilon} + \text{h.c.}\right),
\end{align}
where $g_{k}$ is the vacuum field strength
and $a_{\vec{k} \uvec{\varepsilon}}$ the photon annihilation operator
with wave vector $\vec{k} = \omega_{k} \uvec{k} / c$
and polarization $\uvec{\varepsilon} \perp \vec{k}$.
In the long-time limit,
the change of the electron's spin
can be described by a completely positive map
that is constructed from matrix elements
\begin{align}
  \bracket[\big]{\vec{K}', m_{s}' {}^{+}; m' {\vec{k} \uvec{\epsilon}}}{S}{n \mathrm{S} \tfrac{1}{2}, m_{j}; m {\vec{k} \uvec{\epsilon}}}
  \label{eq:scatmat}
\end{align}
of the scattering operator
\begin{align}
  S & = 1 - 2 \pi \ii \int \mathrm{d}E \delta \left(E - H_{0}\right) T \left(E + \ii 0\right) \delta \left(E - H_{0}\right),
\end{align}
where $\ket{m {\vec{k} \uvec{\varepsilon}}}$
is the $m$-photon number state
of the mode $\vec{k} \uvec{\varepsilon}$
and $H_{0} = H_{\mathcal{A}} + H_{\mathcal{R}}$ the free Hamiltonian
of atom and radiation.
The transition operator $T$
has a perturbative expansion
$T \left(z\right) = V + V \, G_{0} \left(z\right) \, T \left(z\right)$
in powers of $V$ and the unperturbed resolvent
$G_{0} \left(z\right) = \left(z - H_{0}\right)^{-1}$.

\section{Transition Amplitudes}

The relevant transition amplitudes are well known \cite{Yin:1992fu,Yin:1993fk}
and can be written in terms of first and second order
angular, $A_{1}$, $A_{2}$, and radial,
$D_{1}$, $D_{2}$, dipole matrix elements.
For completeness, we present their straightforward derivation.

\begin{figure}
  \includegraphics{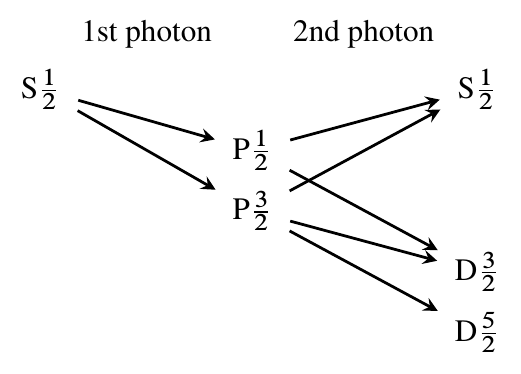}%
  \caption{%
    Electric dipole selection rules between
    $l j$-states in one- and two-photon absorption.
    These two processes are distinguishable by
    the final orbital angular momenta,
    in particular,
    the separation into states with even
    ($\mathrm{S}$, $\mathrm{D}$)
    and those with odd parity ($\mathrm{P}$).%
  }%
  \label{fig:selection_rules}%
\end{figure}

Evaluating the scattering matrix element, Eq.~\eqref{eq:scatmat},
to first order gives the probability amplitude
for the absorption of a single photon
and ejecting an electron with asymptotic wavevector $\vec{K}'$:
$\frac{1}{\hbar} \delta \left(\omega - \omega_{1}\right) \,
\delta_{m, m' + 1} \sqrt{m} \; t_{1} \left(m_{s}', m_{j}\right)$
with $\omega = c k$ and
\begin{multline}
  t_{1} \left(m_{s}', m_{j}\right) = \frac{2 \pi \hbar^{2} g_{k}}{\sqrt{m_{e} K'}} \sum_{m_{l}'} Y_{1 m_{l}'} \bigl(\uvec{K}'\bigr) \smashoperator{\sum_{j' m_{j}'}} \, \clebschgordan*{1 m_{l}'}{\tfrac{1}{2} m_{s}'}{j' m_{j}'} \\ \times \bracket*{m_{j}'}{A_{1} \left(\mathrm{P} j', \mathrm{S} \tfrac{1}{2}; \uvec{\epsilon}\right)}{m_{j}} \, D_{1} \left(E' \mathrm{P} j', n \mathrm{S} \tfrac{1}{2}\right).
  \label{eq:t1}
\end{multline}
The dipole selection rules dictate
that $t_{1}$ only connects to $\mathrm{P}$ continuum states,
$l' = 1$ (see also Fig.~\ref{fig:selection_rules}).
This finds expression
in the angular dipole matrix element,
\begin{widetext}
  \begin{gather}
    \bracket*{m_{j}'}{A_{1} \left(l' j', l j; \uvec{\epsilon}\right)}{m_{j}} = \sqrt{\tfrac{2 l + 1}{2 l' + 1}} \, \clebschgordan*{l 0}{1 0}{l' 0} \smashoperator{\sum_{m_{l}^{\vphantom{\prime}} m_{l}'}} \sum_{q \vphantom{m_{l}'}} \varepsilon_{q} \, \clebschgordan*{l m_{l}^{\vphantom{\prime}}}{1 q}{l' m_{l}'} \sum_{m_{s}} \, \clebschgordan*{l' m_{l}'}{\tfrac{1}{2} m_{s}}{j' m_{j}'} \, \clebschgordan*{l m_{l}}{\tfrac{1}{2} m_{s}}{j m_{j}},
    \label{eq:A1}
  \end{gather}
\end{widetext}
which vanishes unless $l$ and $l'$ differ by one.
Moreover, the projections
$\varepsilon_{q} = \uvec{\varepsilon} \cdot \uvec{e}_{q}^{*}$
of the incident-field polarization onto the spherical basis vectors
$\uvec{e}_{\pm 1} = \mp \left(\uvec{x} \pm \uvec{y}\right) / \sqrt{2}$,
$\uvec{e}_{0} = \uvec{z}$
determine the respective shares of
$\sigma$- ($\Delta m_{j} = \pm 1$)
and $\pi$-transitions ($\Delta m_{j} = 0$)
in the absorption amplitude.
The radial integrals \cite{Fermi:1930qa,Seaton:1951mi},
$D_{1} \left(E' l' j', n l j\right) = e \int_{0}^{\infty} \mathrm{d}x \, x^{3} R_{E' l' j'}^{*} \left(x\right) R_{n l j}^{\vphantom{*}} \left(x\right)$, \etc{},
shall not be explicitly calculated here
and serve as empirical parameters.
In the same manner,
we derive the two-photon transition amplitude
in second order perturbation theory.
It can be written as
$\frac{1}{2 \hbar} \delta \left(\omega - \omega_{2}\right) \, \delta_{m, m' + 2} \sqrt{m \left(m - 1\right)} \; t_{2} \left(m_{s}', m_{j}\right)$,
where we defined
\begin{widetext}
  \begin{gather}
    t_{2} \left(m_{s}', m_{j}\right) = \frac{2 \pi \ii \hbar^{3} g_{k}^{2}}{\sqrt{m_{e} K'}} \smashoperator{\sum_{l' m_{l}'}} Y_{l' m_{l}'} \bigl(\uvec{K}'\bigr) \smashoperator{\sum_{j' m_{j}'}} \, \clebschgordan*{l' m_{l}'}{\tfrac{1}{2} m_{s}'}{j' m_{j}'} \sum_{j''} \, \bracket*{m_{j}'}{A_{2} \left(l' j', \mathrm{P} j'', \mathrm{S} \tfrac{1}{2}; \uvec{\epsilon}\right)}{m_{j}} \, D_{2} \left(E' l' j', \mathrm{P} j'', n \mathrm{S} \tfrac{1}{2}\right),
    \label{eq:t2} \\
    \bracket*{m_{j}'}{A_{2} \left(l' j', l'' j'', l j; \uvec{\epsilon}\right)}{m_{j}} = \sum_{m_{j}''} \, \bracket*{m_{j}'}{A_{1} \left(l' j', l'' j''; \uvec{\epsilon}\right)}{m_{j}''} \, \bracket*{m_{j}''}{A_{1} \left(l'' j'', l j; \uvec{\epsilon}\right)}{m_{j}},
    \label{eq:A2} \\
    \shortintertext{and}
    D_{2} \left(E' l' j', l'' j'', n l j\right) = \sum_{n''} \frac{D_{1} \left(E' l' j', n'' l'' j''\right) D_{1} \left(n'' l'' j'', n l j\right)}{E' - E \left(n'' l'' j''\right) - \hbar c k + \ii 0} + \int \mathrm{d}E'' \, \frac{D_{1} \left(E' l' j', E'' l'' j''\right) D_{1} \left(E'' l'' j'', n l j\right)}{E' - E'' - \hbar c k + \ii 0}.
    \label{eq:D2}
  \end{gather}
\end{widetext}
Inspection of the selection rules reveals that,
for the two photon process,
only partial waves with $l' = 0$ or $2$
contribute to the final amplitude,
\cf{} Fig.~\ref{fig:selection_rules}.

\section{Finite Detection Efficiency}

It can be expected
that an inefficient measurement
of the electron spin polarization
affects the Bell inequality violation
in a qualitatively similar manner
as an inefficient photon threshold measurement.
We therefore concern ourselves
only with the latter.
A photodetector with limited 
photo detection efficiency
can be modelled as
an ideal photodetector
behind an output port
of an unbiased beam splitter.
Let that beam splitter have
the transmissivity 􏲵$\sqrt{\eta}$.
The photodetection with efficiency $\eta$
is adequately described
if we calculate the expectation value
of the CHSH-Bell operator $B$
with respect to the state
\begin{align}
  \ket{f'} & = B_{1o}(\eta) B_{2o}(\eta) B_{1c}(\eta) B_{2c}(\eta) \ket{f},
\end{align}
where the beam splitter operator $B_{1o}$
couples the mode $1o$ to the auxiliary mode $1o'$,
\begin{align}
  B_{1o}(\eta) & = \exp\bigl[\arccos \sqrt{\eta} \bigl(a^{\dagger}_{1o^{\vphantom{\prime}}} a^{\vphantom{\dagger}}_{1o'} - a^{\vphantom{\dagger}}_{1o^{\vphantom{\prime}}} a^{\dagger}_{1o'}\bigr)\bigr].
\end{align}
The operators $B_{2o}$, $B_{1c}$, and $B_{2c}$
are defined analogously.

\section{Numerical Optimization of $\boldsymbol{\abs*{\expval*{B}}}$}

The Bell-CHSH observable $B$
depends on a variety of measurement settings.
For a range of final states $\ket{f'}$,
parametrized by the phase $\phi$
(and the detection efficiency $\eta$),
we want to find the settings for $B$
that allow for the maximum violation
of the Bell-CHSH inequality \eqref{eq:CHSH}.

Introducing the phase $\phi$
(see the main text) is equivalent
to phase-locking the complex amplitudes
$\alpha_{1o}$, $\alpha_{2o}$,
$\alpha_{1c}$, and $\alpha_{2c}$
in the following way:
\begin{subequations}
  \begin{align}
    \Arg \alpha_{1o} & \equiv \phi_{1o} = - \tfrac{\pi}{2} - \kappa_{1} + \kappa_{2o} + 2 \phi_{2c}, \\
    \Arg \alpha_{2o} & \equiv \phi_{2o} = \tfrac{1}{2} \left(- \pi - \kappa_{2o} + \kappa_{2c} + \phi + 2 \phi_{2c}\right), \\
    \Arg \alpha_{1c} & \equiv \phi_{1c} = - \kappa_{1} + \kappa_{2c} - \phi + 2 \phi_{2c}.
  \end{align}
\end{subequations}
Here, we are free to set
$\phi_{2c} \equiv \Arg \alpha_{2c} = 0$,
since the expectation value
of the Bell-CHSH observable
does not depend on this particular phase.

For now, let us also assume that $d_{2} = 0$.
We then have $N_{f_{c}} = 1$ and
\begin{widetext}
  \begin{multline}
    \bracket*{f'}{\Gamma \left(\zeta^{\vphantom{\prime}}\right) \otimes A \left(\vec{\beta}^{\vphantom{\prime}}\right)}{f'}
    = N_{f}^{-2} \Bigl(\bracket{f_{o}}{\Gamma\left(\zeta\right)}{f_{o}} \bracket*{\sqrt{\eta} \alpha_{1o}, \sqrt{\eta} \alpha_{2o}, 0, 0}{A\left(\vec{\beta}\right)}{\sqrt{\eta} \alpha_{1o}, \sqrt{\eta} \alpha_{2o}, 0, 0} \\ + \bracket{f_{o}}{\Gamma\left(\zeta\right)}{f_{c}} \bracket*{\sqrt{\eta} \alpha_{1o}, \sqrt{\eta} \alpha_{2o}, 0, 0}{A\left(\vec{\beta}\right)}{0, 0, \sqrt{\eta} \alpha_{1c}, \sqrt{\eta} \alpha_{2c}} \ee^{-\left(1 - \eta\right) \norm{\vec{\alpha}}^{2} / 2} \\ + \bracket{f_{c}}{\Gamma\left(\zeta\right)}{f_{o}} \bracket*{0, 0, \sqrt{\eta} \alpha_{1c}, \sqrt{\eta} \alpha_{2c}}{A\left(\vec{\beta}\right)}{\sqrt{\eta} \alpha_{1o}, \sqrt{\eta} \alpha_{2o}, 0, 0} \ee^{-\left(1 - \eta\right) \norm{\vec{\alpha}}^{2} / 2} \\ + \bracket{f_{c}}{\Gamma\left(\zeta\right)}{f_{c}} \bracket*{0, 0, \sqrt{\eta} \alpha_{1c}, \sqrt{\eta} \alpha_{2c}}{A\left(\vec{\beta}\right)}{0, 0, \sqrt{\eta} \alpha_{1c}, \sqrt{\eta} \alpha_{2c}}\Bigr)
  \end{multline}
  with $N_{f}^{2} = 2 + \tfrac{1}{\sqrt{2}} \left[\sin \phi - \cos \phi - \cos\left(2 \phi\right)\right] \, \ee^{- \norm{\vec{\alpha}}^{2} / 2}$,
  \begin{subequations}
    \begin{align}
      \bracket{f_{o}}{\Gamma\left(\zeta\right)}{f_{o}} & = - \sin\left(2 \abs{\zeta}\right) \sin\left(\phi + \Arg \zeta\right), \\
      \bracket{f_{o}}{\Gamma\left(\zeta\right)}{f_{c}} & = \tfrac{1}{\sqrt{2}} \exp\left[-\ii \left(\tfrac{\pi}{4} + \phi\right)\right] \left[\cos\left(2 \abs{\zeta}\right) \left(\tfrac{\ii}{\sqrt{2}} + \sin\left(\tfrac{\pi}{4} + \phi\right)\right) + \sin\left(2 \abs{\zeta}\right) \left(-\ii \sin\left(\tfrac{\pi}{4} + \phi\right) + \sin\left(\tfrac{\pi}{4} + \phi + \Arg \zeta\right)\right)\right], \\
      \bracket{f_{c}}{\Gamma\left(\zeta\right)}{f_{o}} & = \tfrac{1}{\sqrt{2}} \exp\left[\ii \left(\tfrac{\pi}{4} + \phi\right)\right] \left[\cos\left(2 \abs{\zeta}\right) \left(-\tfrac{\ii}{\sqrt{2}} + \sin\left(\tfrac{\pi}{4} + \phi\right)\right) + \sin\left(2 \abs{\zeta}\right) \left(\ii \sin\left(\tfrac{\pi}{4} + \phi\right) + \sin\left(\tfrac{\pi}{4} + \phi + \Arg \zeta\right)\right)\right], \\
      \bracket{f_{c}}{\Gamma\left(\zeta\right)}{f_{c}} & = - \cos\left(2 \abs{\zeta}\right) \cos \phi + \sin\left(2 \abs{\zeta}\right) \sin\left(\Arg \zeta\right) \sin \phi,
      \shortintertext{and}
      \bracket*{\vec{\alpha}}{A\left(\vec{\beta}\right)}{\vec{\alpha'}} & = \ee^{- \left(\norm{\vec{\alpha}}^{2} + \norm{\vec{\alpha'}}^{2}\right) / 2} \, \left(2 \ee^{- \norm{\vec{\beta}}^{2} - \vec{\alpha}^{*} \cdot \vec{\beta} - \vec{\alpha}' \cdot \vec{\beta}^{*}} - \ee^{\vec{\alpha}^{*} \cdot \vec{\alpha}'}\right).
    \end{align}
  \end{subequations}
\end{widetext}
From these expressions it becomes clear
that the Bell-CHSH expectation value
$\bracket*{f'}{B}{f'} = \bracket*{f'}{\Gamma \left(\zeta^{\vphantom{\prime}}\right) \otimes A \left(\vec{\beta}^{\vphantom{\prime}}\right)}{f'} + \bracket*{f'}{\Gamma \left(\zeta^{\prime}\right) \otimes A \left(\vec{\beta}^{\vphantom{\prime}}\right)}{f'} + \bracket*{f'}{\Gamma \left(\zeta^{\vphantom{\prime}}\right) \otimes A \left(\vec{\beta}^{\prime}\right)}{f'} - \bracket*{f'}{\Gamma \left(\zeta^{\prime}\right) \otimes A \left(\vec{\beta}^{\prime}\right)}{f'}$
does not depend on the material phases
$\kappa_{1}$, $\kappa_{2o}$, and $\kappa_{2c}$,
nor on the laser phases $\Arg \alpha_{l}$---%
provided that the phases of
the local oscillator amplitudes
$\vec{\beta}$ and $\vec{\beta'}$
are measured in reference to $\Arg \alpha_{l}$,
\ie{}
\begin{align}
  \beta_{l} & = \abs{\beta_{l}} \frac{\alpha_{l}}{\abs{\alpha_{l}}}  \ee^{\ii \Delta_{l}}, & \beta_{l}' & = \abs{\beta_{l}'} \frac{\alpha_{l}}{\abs{\alpha_{l}}} \ee^{\ii \Delta_{l}'},
  \label{eq:betalbetalprime}
\end{align}
where $l = 1o$, $2o$, $1c$, $2c$.

At this point,
the following optimization variables
can be identified:
$\zeta$, $\zeta'$,
$\abs*{\beta_{l}}$, $\Delta_{l}$,
$\abs*{\beta_{l}'}$, and $\Delta_{l}'$.
Fixed parameters
are $\phi$, $\eta$, $d_{2}$,
and the absolute laser amplitudes $\abs*{\alpha_{l}}$.
According to the definitions \eqref{eq:alphao}
and \eqref{eq:alphac} in the main text,
the absolute amplitudes $\abs*{\alpha_{l}}$
are functions
of certain empirical radial integrals
$D_{1}$ and $D_{2}$,
the values of which are not known at this point.
In order to proceed with the optimization,
we sample positive random values for
\begin{align}
  \frac{\abs*{\alpha_{1o}}}{\lambda}, && \frac{\abs*{\alpha_{2o}}}{\sqrt{\lambda}}, && \frac{\abs*{\alpha_{1c}}}{\lambda}, && \frac{\abs*{\alpha_{2c}}}{\sqrt{\lambda}}
  \label{eq:alphallambda}
\end{align}
and add the numerical scaling factor $\lambda$ to
the set of optimization variables.
Globally optimal solutions
have been acquired numerically
by means of the
covariance matrix adaptation evolution strategy
(CMA-ES) \cite{Hansen:2001fu}.
The numerical optimizations have been repeated
for different samples of the absolute amplitudes.
This has shown that the
maximum violation of Ineq.~\eqref{eq:CHSH}
is independent of the particular random
choice of the parameters \eqref{eq:alphallambda}.
The results are depicted in
Fig.~\ref{fig:violation} in the main text.
It is a plot of
the maximally attainable value of
$\abs*{\bracket*{f'}{B}{f'}}$
as a function of $\phi$
and for different detection efficiencies $\eta$.

The optimal values for
$\lambda$, $\abs*{\beta_{l}}$, and $\abs*{\beta_{l}'}$,
$l = 1o$, $2o$, $1c$, $2c$,
depend nontrivially on $\phi$, $\eta$, and
the parameters \eqref{eq:alphallambda}.
These values will therefore not be explicitly discussed.
In contrast,
the optimal values for
$\zeta$, $\zeta'$, $\Delta_{l}$, and $\Delta_{l}'$
depend only on $\phi$ and $\eta$
and are piecewisely continuous in $\phi$.
Furthermore,
the phases $\Delta_{l}$, $\Delta_{l}'$ always fulfill
\begin{subequations}
  \begin{align}
    \Delta_{1o} & = \Delta_{2 o}, & \Delta_{1c} & = \Delta_{2c}, \\
    \Delta_{1o}' & = \Delta_{2 o}', & \Delta_{1c}' & = \Delta_{2c}'.
  \end{align}
\end{subequations}

The case $d_{2} \neq 0$
can be dealt with in a similar fashion
as the case $d_{2} = 0$ above.
While we have not studied it systematically,
we have verified numerically that,
for several randomly chosen,
non-perturbative complex values of $d_{2}$,
the Bell-CHSH inequality
can be violated to a substantial degree.
Qualitatively, the effect of $d_{2}$
is to shift and distort
the plot in Fig.~\ref{fig:violation},
such that the maximum violation 
no longer occurs for $\phi = 0$.

\bibliography{document_revised}

\end{document}